\documentclass[aps,prd,preprint,superscriptaddress,amsmath,amssymb,showpacs]{revtex4-1}
\usepackage{dcolumn}
\usepackage{graphicx}
\usepackage{float}
\usepackage{physics}
\usepackage[colorlinks=true,allcolors=blue]{hyperref}
\UseRawInputEncoding

\begin{document}

\title{Effective running coupling constant and jet quenching parameter in the spinning background from holography}

\author{Zhou-Run Zhu }
\email{zhuzhourun@zknu.edu.cn}
\affiliation{School of Physics and Telecommunications Engineering, Zhoukou Normal University, Zhoukou 466001, China}

\author{Sheng Wang}
\email{shengwang@mails.ccnu.edu.cn}
\affiliation{Institute of Particle Physics and Key Laboratory of Quark and Lepton Physics (MOS), Central China Normal University, Wuhan 430079, China}

\author{Man-Li Tian}
\email{20252018@zknu.edu.cn}
\affiliation{University Clinic, Zhoukou Normal University, Zhoukou 466001, China}

\author{Defu Hou}
\email{houdf@mail.ccnu.edu.cn}
\affiliation{Institute of Particle Physics and Key Laboratory of Quark and Lepton Physics (MOS), Central China Normal University, Wuhan 430079, China}

\begin{abstract}
In this work, we study the effective running coupling constant of heavy quark pair and jet quenching parameter in the spinning background. Ultra-locally, the boosted fluid is described by the boosted parameter and dual to a globally rotating system. Our results show that the angular momentum suppresses the effective running coupling constant and reduces its maximum value. The results demonstrate that the angular momentum promotes the dissociation of quarkonium and has a stronger effect on the effective running coupling constant when the axis of $Q\overline{Q}$ is transverse to the direction of the angular momentum. We also find that the angular momentum enhances the jet quenching parameter and has a stronger effect when the jet moves transversely to the direction of the angular momentum, namely $\hat{q}_{\perp}> \hat{q}_{\parallel}$. We discuss the dependence of the jet quenching parameter on the $\eta/s$ at strong coupling in the presence of the angular momentum.

\end{abstract}
\maketitle

\section{Introduction}\label{sec:01_intro}

Extensive experimental results from heavy-ion collisions have established compelling evidence for quark-gluon plasma (QGP) formation under extreme conditions \cite{Arsene:2004fa,Adcox:2004mh,Back:2004je,Adams:2005dq}. The observed suppression of heavy quarkonium serves as one critical signature for QGP generation \cite{Matsui:1986dk}. The melting of heavy quarkonium provides important insights into the non-equilibrium dynamics of the QGP. The effective running coupling constant describing the interaction between a heavy quark and antiquark is fundamental to understanding the quarkonium dissociation. Researchers have extensively studied the effective running coupling constant in different theories \cite{Giunti:1991ta,Lombardo:1996gp,Aguilar:2001zy,Bloch:2003sk,Kaczmarek:2004gv,Kaczmarek:2005ui,Baikov:2016tgj,Deur:2016tte,Takaura:2018vcy}. Another important characteristic of the QGP is jet quenching, denoting the energy loss undergone by energetic partons passing through the hot, dense medium. Quantitatively, jet quenching parameter $\hat{q}$ is used to characterize this energy loss \cite{Wang:1991xy,Baier:1996kr,Baier:1996sk,Guo:2000nz,Gyulassy:2000er}.

The experimental results reveal that the quark-gluon plasma (QGP) generated in heavy-ion collisions behaves as a strongly coupled fluid \cite{Kaczmarek:2005ui}. AdS/CFT correspondence \cite{Maldacena:1997re,Witten:1998qj,Gubser:1998bc} provides a non-perturbative framework for probing strongly-coupled systems. The quark-antiquark pairs correspond to U-shaped strings connecting boundary endpoints from holography. The effects of chemical potential and magnetic field on the effective running coupling constant within the holographic model have been discussed in \cite{Chen:2021gop}. Recent studies explore the effective running coupling constant in rotating backgrounds \cite{Zhou:2023qtr} and flavor-dependent systems \cite{Guo:2024qiq}. Seminal work by Liu et al. \cite{Liu:2006ug} investigated the jet quenching parameter $\hat{q}$ in $\mathcal{N}$=4 SYM plasma within the holographic framework. The jet quenching parameter is derived from the minimal world-sheet surface bounded by a light-like Wilson loop. Motivated by \cite{Liu:2006ug}, significant research has extended the investigation of the jet quenching parameter in different holographic models. The effect of finite density on $\hat{q}$ has been studied in \cite{Lin:2006au,Armesto:2006zv,Avramis:2006ip}. The jet quenching parameter in the magnetized background has been discussed in \cite{Li:2016bbh,Zhang:2018pyr,Rougemont:2020had,Zhu:2019ujc,Zhu:2023aaq}. Moreover, the jet quenching parameter in Kerr-$AdS_5$ black hole background has been investigated in \cite{Golubtsova:2022ldm}. Other related analyses can see \cite{BitaghsirFadafan:2010rmb,Caceres:2006as,Nakano:2006js,Zhang:2012jd,Li:2014dsa}.

In this work, we explore the effective running coupling constant of heavy quark pair and jet quenching parameter in the spinning background from holography \cite{Hawking:1998kw,Gibbons:2004ai,Gibbons:2004js,Garbiso:2020puw,Amano:2023bhg}. The motivation is from the generation of nonzero angular momentum in non-central heavy-ion collisions \cite{Liang:2004ph,Becattini:2007sr,Baznat:2013zx,STAR:2017ckg,Jiang:2016woz}. The angular momentum induces relativistic rotation of QGP, motivating the effect of rotation on the effective running coupling constant and jet quenching parameter. We consider the Myers-Perry black holes background, in which the compact boundary of the spacetime supports a gauge theory defined on $S^3 \times \mathbb{R}$ \cite{Garbiso:2020puw}. Myers-Perry black holes are solutions to the vacuum Einstein equations in higher dimensions that describe rotating black holes. The connection between these objects and QCD is not a direct physical one, but rather a holographic duality​. In the framework of AdS/CFT correspondence, the Myers-Perry black hole can be reduced to a Schwarzschild black brane boosted along the $x_3$ direction with boost parameter $a$ in the large black hole (high temperature) limit. This boost parameter is directly related to the vorticity of the dual fluid. The relation is motivated by the need to understand strongly coupled systems with rotation. Myers-Perry black holes serve as a theoretical laboratory​ where calculations are tractable on the gravity side, providing insights into the complex, non-perturbative dynamics of rotating quark-gluon plasmas in QCD. Projecting this fluid flow onto flat spacetime presents a fluid with non-trivial vorticity and expansion profiles. In order to analyze the effect of rotation, one can consider zooming in on a small patch and find that the fluid exhibits uniform streaming motion. This ultra-local description corresponds to a globally rotating fluid, with the planar black brane providing the essential holographic framework for studying QGP dynamics in the dual field theory. It should be mentioned that the heavy quark dynamics in the rotating background have been explored in \cite{Arefeva:2020jvo,Golubtsova:2021agl}. The thermodynamics of heavy quarkonium in the spinning black hole have been studied in \cite{Zhu:2024uwu,Zhu:2024dwx,Zhu:2025ucq}.

The paper is organized as follows. In Sec.~\ref{sec:02}, we briefly review the Myers-Perry black hole background. In Sec.~\ref{sec:03}, we discuss the effective running coupling constant of heavy quark pair and jet quenching parameter in the spinning background. In Sec.~\ref{sec:04}, we give the conclusion and discussion.

\section{Spinning Myers-Perry black hole background}\label{sec:02}

We first briefly review the spinning black hole background studied by Hawking et al. The metric is given by \cite{Hawking:1998kw}
\begin{equation}
\label{eqc1}
\begin{split}
ds^{2} & =-\frac{\Delta}{\rho^{2}}(dt_{H}-\frac{a\sin^{2}\theta_{H}}{\Xi_{a}}d\phi_{H}-\frac{b\cos^{2}\theta_{H}}{\Xi_{b}}d\psi_{H})^{2}\\
 & +\frac{\Delta_{\theta_{H}}\sin^{2}\theta_{H}}{\rho^{2}}(adt_{H}-\frac{r_{H}^{2}+a^{2}}{\Xi_{a}}d\phi_{H})^{2}+\frac{\Delta_{\theta_{H}}\cos^{2}\theta_{H}}{\rho^{2}}(bdt_{H}-\frac{r_{H}^{2}+b^{2}}{\Xi_{b}}d\psi_{H})^{2}+\frac{\rho^{2}}{\Delta}dr_{H}^{2}\\
 & -\frac{\rho^{2}}{\Delta_{\theta_{H}}}d\theta_{H}^{2}+\frac{1+\frac{r_{H}^{2}}{R^{2}}}{r_{H}^{2}\rho^{2}}(abdt_{H}-\frac{b\left(r^{2}+a^{2}\right)\sin^{2}\theta_{H}}{\Xi_{a}}d\phi_{H}-\frac{a\left(r^{2}+b^{2}\right)\cos^{2}\theta_{H}}{\Xi_{b}}d\psi_{H})^{2},
 \end{split}
\end{equation}
with
 \begin{equation}
\label{eqc11}
\begin{split}
\Delta=\frac{1}{r_{H}^{2}}(r_{H}^{2}+a^{2})(r_{H}^{2}+b^{2})(1+\frac{r_{H}^{2}}{R^{2}})-2M,\\
 \Delta_{\theta_{H}}=1-\frac{a^{2}}{R^{2}}\cos^{2}\theta_{H}-\frac{b^{2}}{R^{2}}\sin^{2}\theta_{H},\\
 \rho=r_{H}^{2}+a^{2}\cos^{2}\theta_{H}+b^{2}\sin^{2}\theta_{H},\\
 \Xi_{a}=1-\frac{a^{2}}{R^{2}},\\
\Xi_{b}=1-\frac{b^{2}}{R^{2}},\\
 \end{split}
\end{equation}
where $\phi_H$, $\psi_H$, and $\theta_H$ represent Hopf angular coordinates. $t_H$, $R$, and $r_H$ represent time, AdS radius, and radial coordinate, respectively. $a$, $b$ are angular momentum parameters. We consider the $a = b$ case, corresponding to the spinning Myers-Perry black hole \cite{Gibbons:2004ai,Gibbons:2004js}.

One can simplify the calculations by using the following coordinates \cite{Murata:2008xr}
 \begin{equation}
\label{eqc111}
\begin{split}
t=t_{H},\\
 r^{2}=\frac{a^{2}+r_{H}^{2}}{1-\frac{a^{2}}{R^{2}}},\\
\theta=2\theta_{H},\\
 \phi=\phi_{H}-\psi_{H},\\
\psi=-\frac{2at_{H}}{R^{2}}+\phi_{H}+\psi_{H},\\
b=a,\\
\mu=\frac{M}{(R^{2}-a^{2})^{3}}.\\
 \end{split}
\end{equation}

The metric (\ref{eqc1}) can be rewritten as
 \begin{equation}
\label{eqc2}
\begin{split}
ds^{2}=-(1+\frac{r^{2}}{R^{2}})dt^{2}+\frac{dt^{2}}{G(\text{r)}}+\frac{r^{2}}{4}((\sigma^{1})^{2}+(\sigma^{2})^{2}+(\sigma^{3})^{2})+\frac{2\mu}{r^{2}}(dt+\frac{a}{2}\sigma^{3})^{2},
 \end{split}
\end{equation}
with
 \begin{equation}
\label{eqc3}
\begin{split}
G(r)=1+\frac{r^{2}}{R^{2}}-\frac{2\mu(1-\frac{a^{2}}{R^{2}})}{r^{2}}+\frac{2\mu a^{2}}{r^{4}},\\
\mu=\frac{r_{h}^{4}(R^{2}+r_{h}^{2})}{2R^{2}r_{h}^{2}-2a^{2}(R^{2}+r_{h}^{2})},\\
\sigma^{1}=-sin\psi dtd\theta+cos\psi sin\theta d\phi,\\
\sigma^{2}=cos\psi d\theta+sin\psi sin\theta d\phi,\\
\sigma^{3}=d\psi+cos\theta d\phi,\\
 \end{split}
\end{equation}
where
\begin{equation}
\label{eqc4}
\begin{split}
-\infty<t<\infty,\ r_{h}<r<\infty,\ 0\leq\theta\leq\pi,\ 0\leq\phi\leq2\pi,\ 0\leq\psi\leq4\pi.
 \end{split}
\end{equation}

Applying a coordinate transformation yields the planar black brane \cite{Garbiso:2020puw}
 \begin{equation}
\label{eqc5}
\begin{split}
t=\tau,\\
\frac{R}{2}(\phi-\pi)=x_{1},\\
\frac{R}{2}tan(\theta-\frac{\pi}{2})=x_{2},\\
\frac{R}{2}(\psi-2\pi)=x_{3},\\
r=\tilde{r},\\
 \end{split}
\end{equation}
where $(\tau, \widetilde{r}, x_1, x_2, x_3)$ denote the new coordinates. One can rescale these coordinates by $\beta$
\begin{equation}
\label{eqc6}
\begin{split}
\tau\rightarrow\beta^{-1}\tau,\\
x_{1}\rightarrow\beta^{-1}x_{1},\\
x_{2}\rightarrow\beta^{-1}x_{2},\\
x_{3}\rightarrow\beta^{-1}x_{3},\\
\tilde{r}\rightarrow\beta\tilde{r},\\
\tilde{r_{h}}\rightarrow\beta\tilde{r_{h}}.(\beta\rightarrow \infty)\\
 \end{split}
\end{equation}

The boosted Schwarzschild black brane metric in the $\tau$-$x_3$ plane is given by
\cite{Garbiso:2020puw}
\begin{equation}
\label{eqc7}
\begin{split}
ds^{2}=\frac{r^{2}}{R^{2}}(-d\tau^{2}+dx_{1}^{2}+dx_{2}^{2}+dx_{3}^{2}+\frac{r_{h}^{4}}{r^{4}(1-\frac{a^{2}}{R^{2}})}(d\tau+\frac{a}{R}dx_{3})^{2})+\frac{R^{2}r^{2}}{r^{4}-r_{h}^{4}}dr^{2},
 \end{split}
\end{equation}
where $a$ is the boost parameter. The Schwarzschild black brane is recovered when parameter $a=0$.

We employ $z$ ($r=1/z$) as the holographic fifth coordinate.
\begin{equation}
\label{eqc8}
\begin{split}
ds^{2}=\frac{1}{z^{2}R^{2}}[-d\tau^{2}+dx_{1}^{2}+dx_{2}^{2}+dx_{3}^{2}+\frac{z^{4}}{z_{h}^{4}(1-\frac{a^{2}}{R^{2}})}(d\tau+\frac{a}{R}dx_{3})^{2}]+\frac{R^{2}z_{h}^{4}}{z^{2}(z_{h}^{4}-z^{4})}dz^{2}.
 \end{split}
\end{equation}

The expression of temperature is \cite{Garbiso:2020puw}
\begin{equation}
\label{eqb1}
T=\frac{\sqrt{1-a^{2}}}{z_{h}\pi  R^{3}},
\end{equation}
where $z_h$ denotes the horizon. It should be mentioned that the angular velocity $\Omega = a/R^2$ in the limitation of the large black hole\cite{Garbiso:2020puw}.

\section{Effective running coupling constant and jet quenching parameter in the spinning black hole background}\label{sec:03}

\subsection{Effective running coupling constant in the spinning black hole background}

In this subsection, we explore the effective running coupling constant of heavy quark pairs in the spinning Myers-Perry black holes background. The Nambu-Goto action of $Q\overline{Q}$ is
\begin{equation}
\label{eq10}
S= -\frac{1}{2\pi\alpha'} \int d\tau d\sigma \sqrt{-det g_{\alpha \beta}},
\end{equation}
where $g_{\alpha \beta}$ represent the determinant of induced metric.

The metric (\ref{eqc8}) implies rotation about the $x_3$ axis. Therefore, we examine the effective running coupling constant in the transverse and parallel cases. The $Q\overline{Q}$ pair axis lies in the $x_1$-$x_2$ plane for the transverse case, and along the $x_3$ axis for parallel case.

We discuss the transverse case first. The coordinates are parameterized by
\begin{equation}
\label{eq11}
\tau=\xi,\ x_{1}=\eta,\ x_{2}=0,\ x_{3}=0,\ z=z(\eta),
\end{equation}

The expression of Lagrangian density is
\begin{equation}
\label{eq12}
\mathcal{L}=\sqrt{A(z)+B(z)\dot{z}^{2}},
\end{equation}
with
 \begin{equation}
\label{eq13}
\begin{split}
A(z)=A(z_\perp)=\frac{1}{z^{4}R^{4}}-\frac{1}{R^{4}z_{h}^{4}(1-\frac{a^{2}}{R^{2}})},\
B(z)=B(z_\perp)=\frac{z_{h}^{4}}{z^{4}(z_{h}^{4}-z^{4})}-\frac{1}{(z_{h}^{4}-z^{4})(1-\frac{a^{2}}{R^{2}})}.
 \end{split}
\end{equation}

For the parallel case, one can parameterize coordinates using
\begin{equation}
\label{eq111}
\tau=\xi,\ x_{1}=0,\ x_{2}=0,\ x_{3}=\eta,\ z=z(\eta).
\end{equation}

The expressions of $A(z)$ and $B(z)$ are
 \begin{equation}
\label{eq13}
\begin{split}
A(z)=A(z_\parallel)=\frac{z_{h}^{4}-z^{4}}{z^{4}z_{h}^{4}R^{4}},\
B(z)=B(z_\parallel)=\frac{z_{h}^{4}}{z^{4}(z_{h}^{4}-z^{4})}-\frac{1}{(z_{h}^{4}-z^{4})(1-\frac{a^{2}}{R^{2}})}.
 \end{split}
\end{equation}

The interquark distance $L$ of $Q\overline{Q}$ is
 \begin{equation}
\label{eq14}
\begin{split}
L=2\int_{0}^{z_{c}}\sqrt{\frac{A(z_{c})B(z)}{A(z)^{2}-A(z)A(z_{c})}}dz.
 \end{split}
\end{equation}

The free energy of $Q\overline{Q}$ is
 \begin{equation}
\label{eq15}
\begin{split}
\frac{\pi F_{Q\overline{Q}}}{\sqrt{\lambda}}=\int_{0}^{z_{c}}(\sqrt{\frac{A(z)B(z)}{A(z)-A(z_{c})}}-\sqrt{A(z \rightarrow0)})dz-\int_{z_{c}}^{\infty} \sqrt{A(z \rightarrow0)}dz,
 \end{split}
\end{equation}
where $\sqrt{\lambda}=\frac{L^{2}}{\alpha'}$ denote the 't Hooft coupling. We take $\lambda=1$ in the following calculations. To eliminate free energy divergences, one can use minimum subtraction \cite{Ewerz:2016zsx}.

The free energy of a single quark is
 \begin{equation}
\label{eq17}
\begin{split}
\frac{F_{Q}}{\sqrt{\lambda}}=\frac{1}{2\pi}[\int_{0}^{z_{h}}(\sqrt{B(z)}-\frac{1}{z^{2}})dz-\frac{1}{z_{h}}].
 \end{split}
\end{equation}

The the potential energy $V_{Q\overline{Q}}$ is $V_{Q\overline{Q}}=F_{Q\overline{Q}}-2F_{Q}$ \cite{Ewerz:2016zsx}.

The expression of effective running coupling constant of quark-antiquark pair can be obtained from lattice QCD \cite{Kaczmarek:2005ui}
 \begin{equation}
\label{eq151}
\begin{split}
\alpha=\frac{3L^2}{4}\frac{d V_{Q\overline{Q}}}{d L}.
 \end{split}
\end{equation}

The temperature dependence of $\alpha$ at short and large distances has been studied from lattice QCD \cite{Kaczmarek:2005ui}. In the short-distance limit, $\alpha$ exhibits the weakening with decreasing distance that is typical of the perturbative regime. At large distances, the behavior of the coupling is suppressed due to color screening. As a consequence of these two opposing trends, $\alpha$ reaches a maximum at an intermediate distance scale. It is important to note that while color screening suppresses the coupling in the large distance regime, non-perturbative effects still play a dominant role in determining its magnitude \cite{Kaczmarek:2005ui}.

\begin{figure}[H]
    \centering
      \setlength{\abovecaptionskip}{0.1cm}
    \includegraphics[width=9cm]{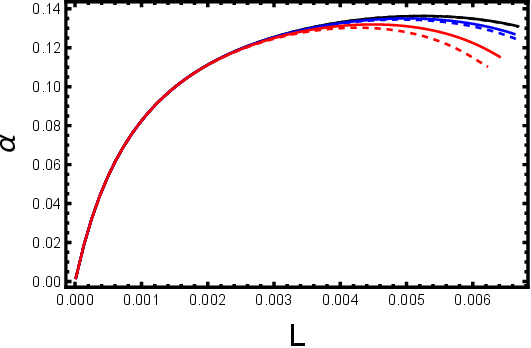}
    \caption{\label{fig1} Effective running coupling constant of $Q\overline{Q}$ versus interquark distance $L$ with different boost parameter $a$. From top to bottom $a = 0.1,\ 0.2,\ 0.3$, respectively. The solid line (dashed line) represents the parallel (transverse) case.}
\end{figure}

We study the effect of the boost parameter on the effective running coupling constant of $Q\overline{Q}$ in the spinning black hole background, which is dual to a rigidly rotating fluid. In the following calculations, we set $R=1$. Following \cite{Garbiso:2020puw}, we consider the planar black brane and take $T=100/\pi$, and take the boost parameter within the stable regime $a < 0.75R$.

In Fig.~\ref{fig1}, we depict the effective running coupling constant of $Q\overline{Q}$ versus interquark distance $L$ with different boost parameter $a$. The effective coupling constant characterizes the interaction strength between the quark-antiquark pair in the medium. A large coupling indicates strong interactions where quark interactions dominate. However, a small coupling implies quarks behave nearly independently. We observed that the effective coupling constant initially increases with interquark distance $L$ until it reaches a maximum value. It implies that the interaction force between quarks intensifies until reaching a maximum value, after which the interaction of heavy quark-antiquark pair becomes weak. Moreover, one can find that the boost parameter suppresses the effective running coupling constant and reduces its maximum value. From the results, the boost parameter promotes the dissociation of quarkonium. The angular velocity $\Omega = a/R^2$\cite{Garbiso:2020puw}. Therefore, the angular momentum promotes the dissociation of quarkonium.

This result can be explained from the potential energy of heavy quarkonium. The authors of Ref. \cite{Zhu:2024dwx} investigated the potential energy of heavy quarkonium in this background and found that it rapidly approaches zero as the angular momentum increases. This indicates that heavy quarkonium becomes more unstable under stronger rotation. In the same spinning black hole background, the authors of Ref. \cite{Zhu:2024uwu} studied the spectral function of heavy quarkonium and found that an increasing angular momentum facilitates its dissociation. Our results are consistent with the results of Refs. \cite{Zhu:2024dwx,Zhu:2024uwu}. Furthermore, the angular momentum has a stronger effect on the effective running coupling constant when the axis of $Q\overline{Q}$ is transverse to the direction of the angular momentum.

\subsection{Jet quenching parameter in the spinning black hole background}

In this subsection, we study the jet quenching parameter $\hat{q}$ in the spinning Myers-Perry black holes background. Following the holographic approach established by Liu et al. \cite{Liu:2006ug}, the jet quenching parameter $\hat{q}$ is extracted from the expectation value of a light-like Wilson loop. In this framework, $\hat{q}$ characterizes the transverse momentum broadening of a fast parton traversing the strongly coupled medium. The jet quenching parameter can be extracted from \cite{Liu:2006ug}
\begin{equation}
<W^A[{\cal C}]> \approx \exp [-\frac{1}{4\sqrt{2}}\hat{q}L^{-}L_k^2],
\label{jet}
\end{equation}
where $<W^A[{\cal C}]>\approx <W^F[{\cal C}]>^2$. $<W^F[{\cal C}]>$ denotes the expectation value in the fundamental representation from AdS/CFT correspondence. One can get\cite{Liu:2006ug}
\begin{equation}
<W^F[{\cal C}]>\approx\exp[-S_I] \label{WF}.
\end{equation}

Therefore, the expression for the jet quenching parameter in $\mathcal{N}$=4 SYM plasma is \cite{Liu:2006ug}
\begin{equation}
\hat{q}=8\sqrt{2}\frac{S_I}{L^{-}L_k^2},\label{q}
\end{equation}
where $S_{I} = S - S_{0}$. $S$ denote the total energy and $S_{0}$ is the self-energy. $L_k$ and $L^{-}$ are the size of the null-like rectangular contour.

Due to the rotation direction being along $x_3$ axis. We can consider the transverse and parallel cases. In the transverse case, the jets propagate transverse to the rotation axis, while momentum broadening could occur both transverse and parallel to the rotation axis. First, the jets propagate along $x_1$ with momentum broadening along $x_2$, characterized by $\hat{q}_{(\perp,\perp)}$. Second, jets propagate along $x_1$ with momentum broadening along $x_3$, characterized by $\hat{q}_{(\perp,\parallel)}$. We first calculate the $\hat{q}_{(\perp,\perp)}$.

By using the light-cone coordinates
\begin{equation}
\label{eq17}
d\tau=\frac{dx^{+}+dx^{-}}{\sqrt{2}}, \ dx_1=\frac{dx^{+}-dx^{-}}{\sqrt{2}}.
\end{equation}

One can rewrite the metric (\ref{eqc8}) as
\begin{equation}
\label{eq18}
ds^{2}=\frac{1}{z^{2}}[dx_{2}^{2}+dx_{3}^{2}-2dx^{+}dx^{-}+\frac{z^4}{z_{h}^{4}(1-a^{2})}(\frac{dx^{+}+dx^{-}}{\sqrt{2}}+a dx_{3})^2]+\frac{z_{h}^{4}}{z^{2}(z_{h}^{4}-z^{4})}dz^{2}.
\end{equation}

The ansatz for the string configuration is
\begin{equation}
\label{eqc19}
\begin{split}
x^{-}=\xi,\ x_{2}=\sigma,\  x^{+}=x_{3}=const,\ z=z(\sigma).
 \end{split}
\end{equation}

Then Eq.(\ref{eq18}) becomes
\begin{equation}
\label{eq20}
ds^{2}=\frac{1}{2}\frac{z^2}{z_{h}^{4}(1-a^{2})}d\xi^2+(\frac{1}{z^{2}}+\frac{z_{h}^{4}}{z^{2}(z_{h}^{4}-z^{4})}\dot{z}^2)d\sigma^2,
\end{equation}
with $\dot{z}=\frac{dz}{d\sigma}$.

The Nambu-Goto action is
\begin{equation}
\label{eq21}
\ S=\frac{\sqrt{2}L^{-}}{2\pi\alpha'} \int^{\frac{L_k}{2}}_{0}d\sigma \sqrt{\frac{1}{1-a^2}(\frac{1}{z_{h}^{4}}+\frac{1}{z_{h}^{4}-z^{4}}\dot{z}^2)}.
\end{equation}

The Lagrangian density is $\sigma$ independent, implying a conserved Hamiltonian, namely $\mathcal{L}-\frac{\partial \mathcal{L}}{\partial \dot{z}}\dot{z}=E$. Then one can get
\begin{equation}
\label{eq221}
\ \dot{z}=\frac{1}{E^2}\sqrt{\frac{z_{h}^{4}-z^{4}}{z_{h}^{4}}}\sqrt{\frac{1}{z_{h}^{4}(1-a^2)}-E^2}.
\end{equation}

In the low energy limit ($E\rightarrow0$), integrating Eq.(\ref{eq221}) to leading order of $E^2$ yields
\begin{equation}
\label{eq22}
\ L_k=2E\int^{z_h}_{0}dz \sqrt{\frac{z_{h}^{8}(1-a^2)}{z_{h}^{4}-z^{4}}}.
\end{equation}

By substituting (\ref{eq221}) into (\ref{eq21}) and expanding to leading order of $E^2$
\begin{equation}
\label{eq23}
\ S=\frac{\sqrt{2}L^{-}}{2\pi\alpha'} \int^{z_h}_{0}dz \sqrt{\frac{1}{(z_{h}^{4}-z^{4})(1-a^2)}}(1+\frac{E^2 z_{h}^{4}(1-a^2)}{2}).
\end{equation}

The self energy is
\begin{equation}
\label{eq24}
\ S_0=\frac{\sqrt{2}L^{-}}{2\pi\alpha'} \int^{z_h}_{0}dz \sqrt{\frac{1}{(z_{h}^{4}-z^{4})(1-a^2)}}.
\end{equation}

The action $S_{I} = S - S_{0}$ is
\begin{equation}
\label{eq25}
\ S_{I}=\frac{\sqrt{2}L^{-}E^2}{4\pi\alpha'} \int^{z_h}_{0}dz \sqrt{\frac{z_{h}^{8}(1-a^2)}{z_{h}^{4}-z^{4}}}.
\end{equation}

Then one can drive the jet quenching parameter as
\begin{equation}
\label{eq26}
\ \hat{q}_{(\perp,\perp)}=\frac{1}{\pi\alpha'}\frac{1}{I_{(\perp,\perp)}},
\end{equation}
where
\begin{equation}
\label{eq27}
\ I_{(\perp,\perp)}=\int^{z_h}_{0}dz \sqrt{\frac{z_{h}^{8}(1-a^2)}{z_{h}^{4}-z^{4}}}.
\end{equation}

Then we calculate the $\hat{q}_{(\perp,\parallel)}$. In this case, we use the following ansatz \begin{equation}
\label{eqc29}
\begin{split}
x^{-}=\xi,\ x_{3}=\sigma,\  x^{+}=x_{2}=const,\ z=z(\sigma).
 \end{split}
\end{equation}

Then Eq.(\ref{eq18}) becomes
\begin{equation}
\label{eq30}
ds^{2}=\frac{1}{2}\frac{z^2}{z_{h}^{4}(1-a^{2})}d\xi^2+(\frac{1}{z^{2}}+\frac{z^2 a^2}{z_{h}^{4}(1-a^{2})}+\frac{z_{h}^{4}}{z^2(z_{h}^{4}-z^{4})}\dot{z}^2)d\sigma^2+\frac{\sqrt{2}a z^2}{z_{h}^{4}(1-a^{2})}d\xi d\sigma.
\end{equation}

The expression of Nambu-Goto action is
\begin{equation}
\label{eq31}
\ S=\frac{\sqrt{2}L^{-}}{2\pi\alpha'} \int^{\frac{L_k}{2}}_{0}d\sigma \sqrt{\frac{1}{1-a^2}(\frac{1}{z_{h}^{4}}+\frac{1}{z_{h}^{4}-z^{4}}\dot{z}^2)}.
\end{equation}

One can find that the Nambu-Goto action of $\hat{q}_{(\perp,\parallel)}$ (Eq.\ref{eq31}) is equal to the Nambu-Goto action of $\hat{q}_{(\perp,\perp)}$ (Eq.\ref{eq21}), which yields $\hat{q}_{(\perp,\parallel)}$=$\hat{q}_{(\perp,\perp)}$. Thus we denote $\hat{q}_\perp \equiv\hat{q}_{(\perp,\parallel)}=\hat{q}_{(\perp,\perp)}$.

In parallel case, the jets propagate along the $x_{3}$ and the momentum broadening along the $x_{1}$, namely $\hat{q}_{(\parallel,\perp)}$. The light-cone coordinates are
\begin{equation}
\label{eq32}
d\tau=\frac{dx^{+}+dx^{-}}{\sqrt{2}}, \ dx_3=\frac{dx^{+}-dx^{-}}{\sqrt{2}}.
\end{equation}

Then one can rewrite the metric (\ref{eqc8})
\begin{equation}
\label{eq33}
ds^{2}=\frac{1}{z^{2}}[dx_{1}^{2}+dx_{2}^{2}- 2dx^{+}dx^{-}+\frac{z^4}{z_{h}^{4}(1-a^{2})}(\frac{dx^{+}+dx^{-}}{\sqrt{2}}+a \frac{dx^{+}-dx^{-}}{\sqrt{2}})^2 ]+\frac{z_{h}^{4}}{z^{2}(z_{h}^{4}-z^{4})}dz^{2}.
\end{equation}

The ansatz can be
\begin{equation}
\label{eqc34}
\begin{split}
x^{-}=\xi,\ x_{1}=\sigma,\  x^{+}=x_{2}=const,\ z=z(\sigma).
 \end{split}
\end{equation}

Then one can obtain Eq.(\ref{eq33}) as
\begin{equation}
\label{eq35}
ds^{2}=\frac{z^2(1-a)}{z_{h}^{4}(1+a)}d\xi^2+(\frac{1}{z^{2}}+\frac{z_{h}^{4}}{z^{2}(z_{h}^{4}-z^{4})}\dot{z}^2)d\sigma^2.
\end{equation}

The jet quenching parameter in parallel case is
\begin{equation}
\label{eq37}
\ \hat{q}_{(\parallel,\perp)}=\frac{1}{\pi\alpha'}\frac{1}{I_{(\parallel,\perp)}},
\end{equation}
where
\begin{equation}
\label{eq38}
\ I_{(\parallel,\perp)}=\int^{z_h}_{0}dz \frac{1}{z_{h}^{4}}\sqrt{\frac{1-a^2}{z_{h}^{4}-z^{4}}}.
\end{equation}

\begin{figure}[H]
    \centering
      \setlength{\abovecaptionskip}{0.1cm}
    \includegraphics[width=9cm]{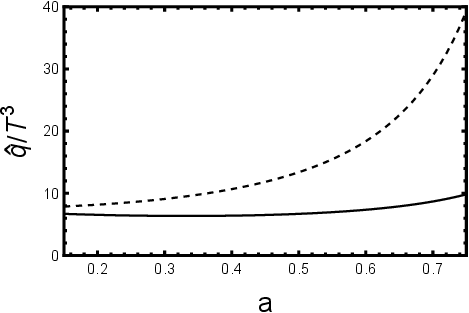}
    \caption{\label{fig2} Jet quenching parameter $\hat{q}/ T^3$ versus boost parameter $a$. The solid line (dashed line) represents the $\hat{q}_\parallel$ ($\hat{q}_\perp$). }
\end{figure}

For simplicity, we denote $\hat{q}_\parallel \equiv \hat{q}_{(\parallel,\perp)}$. Therefore, we study the effect of the boost parameter on $\hat{q}_\perp$ and $\hat{q}_\parallel$. In Fig.~\ref{fig2}, we plot the jet quenching parameter $\hat{q}/ T^3$ versus boost parameter $a$. One can observe that the boost parameter enhances the jet quenching parameter, suggesting the jet will lose more energy in the spinning black hole background. The angular velocity is proportional to the boost parameter \cite{Garbiso:2020puw}. Thus, the angular momentum favors the jet quenching parameter. This finding is consistent with the result in the Kerr-$AdS_5$ black hole background \cite{Golubtsova:2022ldm}. The enhanced energy loss of jets in a rotating medium can be explained by the centrifugal force, which increases their interaction with the medium. This result agrees with Ref. \cite{Chen:2023yug}, where rotation was shown to amplify heavy quark energy loss via centrifugal effects. It should be mentioned that the authors of Ref. \cite{Saha:2019xbl} discuss the jet quenching parameter in the moving plasma. Furthermore, the angular momentum has a slight effect on $\hat{q}_\parallel$, while significantly enhancing $\hat{q}_\perp$. It implies that the jet may lose more energy when moving transverse to the direction of the angular momentum.

We find that the angular momentum enhances the jet quenching parameter and has a stronger effect when the jet moves transversely to the direction of the angular momentum, namely $\hat{q}_{\perp}> \hat{q}_{\parallel}$. These findings reveal an estimation of the jet quenching parameter's dependence on $\eta/s$ at strong coupling. As demonstrated in \cite{Garbiso:2020puw}, $\eta_\perp/s$ maintains unchanged with the angular momentum, whereas $\eta_\parallel/s$ decreases as the angular momentum increases. We refrain from commenting on why only one shear viscosity component saturates the KSS bound while the other violates it which is also observed in the anisotropic backgrounds \cite{Erdmenger:2010xm,Critelli:2014kra,Rebhan:2011vd}. We focus on the $\eta_\parallel/s$, which decreases as the angular momentum increases. This suggests that the fluid is closer to the perfect fluid when the angular momentum grows. We find that the angular momentum favors the jet quenching parameter, suggesting the jet quenching parameter increases as $\eta/s$ decreases in the strongly coupled system.

It should be mentioned that the precise definition of $\hat{q}$ varies depending of the framework and, in the most sound approaches, it is not really a parameter but a function of the transverse momentum. While the jet quenching parameter $\hat{q}$ is a widely used phenomenological​ measure of medium interaction, the reaction rate​  is  more fundamental​ from a theoretical physics perspective, which can be related to the expectation value of light-like Wilson lines.

In heavy-ion collision physics, $\hat{q}$ and the reaction rate are two core quantities that characterize the energy loss of high-energy partons in the QGP, each with distinct advantages and limitations. While $\hat{q}$ offers a more intuitive physical picture and is defined as the mean squared transverse momentum gain per unit path length. The value of $\hat{q}$ can be extracted by analyzing the suppression of high transverse momentum hadron spectra ($R_{AA}$). However, extracting $\hat{q}$ from experimental data depends strongly on assumptions about the jet energy loss mechanism and the medium evolution model. Moreover, in the moderate transverse momentum region or near the critical temperature $T_c$, non-perturbative effects may become significant, limiting the reliability of perturbative QCD calculations. In contrast, the jet reaction rate explicitly gives the probability for a parton to lose energy through a specific process. It is grounded in microscopic interaction cross-sections, offering a solid physical foundation. Nevertheless, compared with $\hat{q}$, the interaction rate is more difficult to extract directly from a single experimental observable and typically requires constraints from complex global fits.

The reaction rate and the jet quenching parameter are not mutually exclusive but complementary physical quantities. Within the holographic framework, both originate from the non-perturbative dynamics of Wilson lines, collectively depicting the overall picture of parton energy loss in strongly coupled plasmas. Our  current work focuses on $\hat{q}$, which serves as a reasonable baseline observable because it is directly linked to experimentally measurable jet quenching phenomena, such as $R_{AA}$. Furthermore, it is worth noting that a potentially significant future work is to explore the synergistic use of both the jet quenching parameter and the reaction rate. This can not only describe jet quenching dynamics more accurately but also serve as versatile probes for revealing deeper static properties, dynamical evolution, and even quantum features of the QGP.

\section{Conclusion and discussion}\label{sec:04}

In this work, we explore the effective running coupling constant of heavy quark pair and jet quenching parameter in the spinning background from holography.

It is found that the effective running coupling constant initially increases with interquark distance $L$ until it reaches a maximum value and then starts to decrease. We also observe that the angular momentum suppresses the effective running coupling constant and reduces its maximum value, indicating the angular momentum promotes the dissociation of quarkonium. Moreover, the angular momentum has a stronger effect on the effective running coupling constant when the axis of $Q\overline{Q}$ is transverse to the direction of the angular momentum. The authors of Ref. \cite{Zhou:2023qtr} study the effective running coupling constant in the local rotating background. They find that angular momentum promotes dissociation and our results agree with Ref. \cite{Zhou:2023qtr}. We also find that the angular momentum enhances the jet quenching parameter and has a stronger effect when the jet moves transversely to the direction of the angular momentum. Although a direct quantitative mapping between the holographic boost parameter $a$ and the angular momentum of the QGP created in heavy-ion collisions is not straightforward, our results demonstrate that within the physically stable regime of the model ($a < 0.75R$), the influence of rotation on the effective coupling strength and the jet quenching parameter is significant and exhibits clear anisotropy. These findings provide qualitative insights into how rotational motion may affect quarkonium dissociation and jet energy loss in a rotating QGP.

The effective coupling $\alpha$ inherently intertwines the running of the coupling and screening effects. One can focus directly on the behavior of potential and cleanly separate vacuum physics from medium-induced effects. It should be mentioned that the potential of quarkonium in the spinning medium has been studied in our previous work (Ref.\cite{Zhu:2024dwx}). We aim to further explore the interaction strength between the quark-antiquark pair and calculate the effective running coupling constant in the spinning medium.

It may be an interesting investigation to discuss the dissociation of quarkonium and jet quenching in the charged Kerr-$AdS_5$ background \cite{Cvetic:2004hs}. We expect explore this work in subsequent studies.

\section*{Acknowledgments}

Defu Hou's research is supported in part by the National Key Research and Development Program of China under Contract No. 2022YFA1604900. Additionally, Defu Hou receives partial support from the National Natural Science Foundation of China (NSFC) under Grant No.12435009, and No. 12275104. Zhou-Run Zhu is supported by the Natural Science Foundation of Henan Province of China under Grant No. 242300420947. Zhou-Run Zhu's work is also supported by the startup Foundation projects for Doctors at Zhoukou Normal University, with the project number ZKNUC2023018.


\end{document}